\documentclass[aps,pra,onecolumn,groupedaddress,showpacs]{revtex4}

\usepackage{amssymb,amsmath}
\usepackage{graphicx}

\newcommand{\om}{\omega}

\newcommand{\pa}{\partial}

\begin{document}

\title{Two-component vector breather solution of the modified BBM  equation}

\author{G. T. Adamashvili}
\affiliation{Georgian Technical University, Kostava str.77, Tbilisi, 0179, Georgia.\\ email: $guram_{-}adamashvili@ymail.com.$ }

\begin{abstract}
This is a continuation of Ref.[1](arXiv:nlin.PS/2001.07758v1). In the present paper, we consider the solution to the modified Benjamin-Bona-Mahony  equation $u_{ t} + C u_{z} + \beta u_{zzt} + a u^{2} u_{z}=0$ using the generalized perturbation reduction method. The equation is transformed to the coupled nonlinear Schr\"odinger equations for auxiliary functions. Explicit analytical expression for the shape and parameters of the two-component vector breather oscillating with the sum and difference of frequencies and wavenumbers are obtained.
\end{abstract}


\maketitle

\section{Introduction}

The nonlinear solitary waves, such as breathers and their various varieties are one of the main objects of study in the nonlinear wave theory. Although these waves are met in completely different physical systems and describe various physical phenomena, their general properties are quite similar. Nonlinear solitary waves can be divided into two main kinds: single-component and two-component solitary waves [1-16].
The properties and the methods of investigations of each of them are completely different. Namely, methods for studying single-component waves are completely unacceptable for studying two-component solitary waves. This is because the study of two-component waves requires a larger number of auxiliary functions and parameters. For instance, the perturbative reduction method (PRM) is adapted for the study of single-component solitary waves, which uses one complex auxiliary function and two constant parameters [17,18]. Later, the generalized version of the PRM is developed in which two complex auxiliary functions and eight constant parameters are used, which made it possible to study two-component nonlinear solitary waves. In the beginning, the generalized PRM was used in nonlinear optics under the condition of self-induced transparency (SIT). Namely, it was proved that the second derivatives in the Maxwell wave equation lead not only to small corrections to the parameters of the SIT pulses, as was previously supposed before applying the generalized PRM [19-27], but also cause the formation of a bound state of two breathers and the formation of a two-component nonlinear wave - vector $0\pi$ pulse. One component of such a vector pulse oscillating with the sum, and the second at the difference of frequencies and wavenumbers (OSDFW). As a result, it was obtained that one of the main SIT pulse is not scalar $0\pi$ pulse, as was previously supposed, but the two-component vector $0\pi$ pulse of SIT, and the scalar $0\pi$  pulse is only some approximation [14-16, 28-35]. Later, a similar two-component vector $0\pi$  pulse was obtained for acoustic nonlinear SIT waves [36-40]. Using the generalized PRM for studying nonresonant nonlinear waves in a dispersive and Kerr-type nonlinear susceptibility medium led to similar results - the formation of a two-component vector pulse OSDFW [41]. A completely different phenomena and corresponding equations, compared with nonlinear equations studied in nonlinear optics and nonlinear acoustics, was considered for surface waves in dispersive nonlinear media, the anharmonic phonons in crystals, acoustic-gravity waves in fluids, in plasma physics, etc. It was proved that in such physical systems also has a solution in the form of a two-component vector breather OSDFW [1].

Even though the generalized PRM was developed relatively recently, using this method, it was possible to establish a connection between a number of the nonlinear differential equations (or system of equations) with the coupled  nonlinear Schr\"odinger equations (NSEs) and find solutions in the form of two-component vector breathers OSDFW. To these equations belong:
the Sin-Gordon equation, the system of Maxwell-Bloch equations, the system of Maxwell-Liouville equations for the linearly and the circularly polarized pulses in an ensemble of semiconductor quantum dots, Maxwell wave equation in the dispersive Kerr-type medium, the system of Maxwell equation and the material equations for two-photon resonant transitions, the system of magnetic Bloch equations and the elastic wave equation,  the modified Korteweg-de Vries equation or the modified Benjamin-Bona-Mahony (BBM) equation [1, 14-16, 28-41].

The purpose of the present work, using the generalized PRM, to consider the modified BBM equation in the form [42-46]
\begin{equation}\label{bbm}
u_{ t}+C u_{z}+\beta u_{zzt} +a u^{2} u_{z}=0,
\end{equation}
where $u(z,t)$ is a real function of space and time, $z$ is spatial variable and $t$ is time variable, $a$, $C$ and $\beta$ are arbitrary constants (see, [47]).

\vskip+0.5cm
\section{The equation for the slowly envelope function}

We can simplify Eq.(1) using the method of slowly changing envelope. For this purpose, we represent  $u$ in the form
\begin{equation}\label{ez}
 u=\sum_{l=\pm1}\hat{u}_{l}Z_{l},\;\;\;\;\;\;\;\;\;\;\;\;Z_{l}= e^{il(kz -\omega t)},
\end{equation}
where $\hat{u}_{l}$ is the slowly varying complex amplitude. This function is complex in view of the fact that the wave is phase modulated.  $Z_{l}= e^{il(k z -\omega t)}$ is the fast oscillating function.To take into account that ${u}$ is a real function, we set $\hat{u}_{1}=\hat{u}_{-1}^{*}$.

The envelope $\hat{u}_{l}$ vary sufficiently slowly in space and time compared with the carrier wave parts, so that the following inequalities are valid
\begin{equation}\label{ap}
 \left|\frac{\partial \hat{u}_{l}}{\partial t}\right|\ll\omega
|\hat{u}_{l}|,\;\;\;\left|\frac{\partial \hat{u}_{l}}{\partial z}\right|\ll k|\hat{u}_{l}|.
\end{equation}

Substituting Eq.(2) in the nonlinear equation (1), and taking into account Eq.(3), we obtain  the dispersion law for the propagating  pulse in the medium
\begin{equation}\label{dis}
\omega=\frac{Ck}{1   -\beta  k^2 }
\end{equation}
and the nonlinear wave equation for the envelope function $\hat{u}_{l}$   in the form:
\begin{equation}\label{equ}
 \sum_{l=\pm 1} Z_{l} [ \frac{\pa \hat{u}_{l}}{\pa t}  +( C -3\beta k^2  )\frac{\pa \hat{u}_{l}}{\pa z}
 + 3\beta ilk \frac{\pa^{2} \hat{u}_{l}}{\pa z^2}  +\beta \frac{\pa^{3} \hat{u}_{l}}{\pa z^3}]
 +a \sum_{L,m,l'} Z_{L+m+l'} \hat{u}_{L} \hat{u}_{m}(il'k  \hat{u}_{l'}+\frac{\pa \hat{u}_{l'}}{\pa z})=0.
\end{equation}

\vskip+0.5cm
\section{Two-component vector breather and the generalized PRM}

For the study of the two-component nonlinear solitary wave solution of Eq.(1) we apply the generalized PRM [1, 14-16, 28-41] by means of which we can transform Eq.(1) into the coupled NSEs. In this method the function $\hat{u}_{l}(z,t)$ can be represented as:
\begin{equation}\label{cemi}
\hat{u}_{l}(z,t)=\sum_{\alpha=1}^{\infty}\sum_{n=-\infty}^{+\infty}\varepsilon^\alpha
Y_{l,n} f_{l,n}^ {(\alpha)}(\zeta_{l,n},\tau),
\end{equation}
where
$$
Y_{l,n}=e^{in(Q_{l,n}z-\Omega_{l,n}
t)},\;\;\;\zeta_{l,n}=\varepsilon Q_{l,n}(z-{v_g}_{l,n}
t),\;\;\;\tau=\varepsilon^2 t,\;\;\;
{v_g}_{l,n}=\frac{d\Omega_{l,n}}{dQ_{l,n}},
$$
$\varepsilon$ is a small parameter. Such an expansion allows us to separate from $\hat{u}_{l}$ the even more slowly changing auxiliary function $ f_{l,n}^{(\alpha )}$. It is assumed that the quantities $\Omega_{l,n}$, $Q_{l,n}$, and $f_{l,n}^{(\alpha)}$ satisfy the conditions:
\begin{equation}\label{ryp}\nonumber\\
\omega\gg \Omega_{l,n},\;\;k\gg Q_{l,n},\;\;\;
\end{equation}
$$
\left|\frac{\partial
f_{l,n}^{(\alpha )}}{
\partial t}\right|\ll \Omega_{l,n} \left|f_{l,n}^{(\alpha )}\right|,\;\;\left|\frac{\partial
f_{l,n}^{(\alpha )}}{\partial z }\right|\ll Q_{l,n}\left|f_{l,n}^{(\alpha )}\right|.
$$
for any value of indexes $l$ and $n$.

Although the quantities  $Q_{l,n}$, $\Omega_{l,n}$, $\zeta_{l,n}$ and ${v_{g;}}_{l,n}$ depend on  $l$ and $n$, for simplicity, we omit these indexes in the next expressions when this does not cause confusion.

The generalized PRM Eq.(6) is acceptable for the  phase modulated complex function $\hat{u}_{l}$. But if the wave is not phase-modulated, then the function $\hat{u}_{l}=\hat{u}_{-l}=\hat{u}$ is real and does not depend on the index $l$.

Substituting the expansion (ansatz) Eq.(6) into Eq.(5) we obtain

\begin{equation}\label{wjh}
 \sum_{l=\pm 1}\sum_{\alpha=1}^{\infty}\sum_{n=-\infty}^{+\infty}\varepsilon^\alpha Z_{l} Y_{l,n} [
\mathcal{W}_{l,n}f_{l,n}^{(\alpha)} +\varepsilon \mathfrak{J}_{l,n}  \frac{\partial f_{l,n}^{(\alpha)}}{\partial
\zeta_{l,n}}  +\mathcal{H}_{l,n} \varepsilon^2 \frac{\partial^{2} f_{l,n}^{(\alpha)}}{\partial
\zeta^{2}_{l,n}}+\mathfrak{h}_{l,n}\varepsilon^2 \frac{\partial f_{l,n}^{(\alpha)}}{\partial \tau} +  O(\varepsilon^3)] $$$$ +a \sum_{L,m,l'} Z_{L+m+l'} \hat{E}_{L} \hat{E}_{m}(il'k  \hat{E}_{l'}+\frac{\pa \hat{E}_{l'}}{\pa z})=0,
 \end{equation}
where
\begin{equation}
\mathcal{W}_{l,n}=-in(\Omega - \beta k^2  \Omega - C Q -2\beta   k \omega  Q  - 2\beta l n k  Q \Omega - l n\beta \omega  Q^{2} - \beta  n^2 Q^{2} \Omega),
$$$$
\mathcal{J}_{l,n}=- Q v_g +\beta k^2   Q v_g   +  C Q + 2\beta   k \omega Q  + 2\beta l n k Q (Q v_g +\Omega) + 2   n l\beta \omega   Q^{2}  + \beta  n^2 Q^{2}(Q v_g +  2 \Omega)],
$$$$
\mathcal{H}_{l,n}= -i \beta Q^{2} [2 (  l k   +  n Q)  v_{g} +  (l \omega  +n \Omega)],
$$$$
\mathfrak{h}_{l,n}= 1-\beta ( l k +  n Q)^{2}.
\end{equation}

Eq.(7) contains four independent equations for different values $l=\pm1$ and $n=\pm1$.

If we equating to zero, the members of the Eq.(7) with the same powers of $\varepsilon$, we will be able to a several of equations. In the first order of $\varepsilon$ we have the equation
\begin{equation}\label{fr}
 \sum_{l=\pm 1}\sum_{n=-\infty}^{+\infty} Z_{l} Y_{l,n} \mathcal{W}_{l,n}\mathfrak{f}_{l,n}^{(1)}=0.
\end{equation}

From Eq.(9) we obtain connection between the parameters $\Omega_{l,n}$ and $Q_{l,n}$:
\begin{equation}\label{diss}
 (\beta k^2-1)  \Omega_{l,n} + (C + 2\beta  k \omega)  Q_{l,n}  +l n 2\beta k Q_{l,n} \Omega_{l,n} +l n \beta \omega  Q^{2}_{l,n} +  \beta   Q^{2}_{l,n} \Omega_{l,n}=0,
\end{equation}
where $l=\pm1$ and $n=\pm1$.

From Eq.(10) we obtain the expression
\begin{equation}\label{rrv}
 {v_{g;}}_{l,n}=\frac{ C + 2\beta ( k \omega   +l n  k \Omega_{l,n} + l n  \omega  Q_{l,n} +   Q_{l,n} \Omega_{l,n})}
 {1- \beta (l k +n  Q_{l,n})^{2}}.
\end{equation}

Eq.(10) consists from the four equations with different values of the indexes $l$ and $n$. But these four equations are reduced to the two independent equations

When $l=\pm1,\;n=\pm1$, than $f_{\pm1,\pm1}^{(1)}\neq0$ and we have the equation
\begin{equation}\label{dis1}
 (\beta k^2-1)  \Omega_{\pm1,\pm1} + (C + 2\beta  k \omega)  Q_{\pm1,\pm1}  + 2\beta k Q_{\pm1,\pm1} \Omega_{\pm1,\pm1} + \beta \omega  Q^{2}_{\pm1,\pm1} +  \beta   Q^{2}_{\pm1,\pm1} \Omega_{\pm1,\pm1}=0,
\end{equation}
and when $l=\pm1,\;n=\mp1$, than $f_{\pm1,\mp1}^{(1)}\neq0$ and we have
\begin{equation}\label{dis2}
 (\beta k^2-1)  \Omega_{\pm1,\mp1} + (C + 2\beta  k \omega)  Q_{\pm1,\mp1}  - 2\beta k Q_{\pm1,\mp1} \Omega_{\pm1,\mp1} - \beta \omega  Q^{2}_{\pm1,\mp1} +  \beta   Q^{2}_{\pm1,\mp1} \Omega_{\pm1,\mp1}=0.
\end{equation}

Taking into account Eqs.(12) and (13), from Eq.(7) in the second order of $\varepsilon$ we obtain
$$
\mathfrak{J}_{\pm1,\pm1}=\mathfrak{J}_{\pm1,\mp1}=0,\;\;\;\;\;\;\;\;
 f_{+1,\pm2}^{(2)}=f_{-1,\pm2}^{(2)}=0.
$$

From the equation (7) we have in the third order of $\varepsilon$ the system of equations proportional $Z_{+1}$ and $Z_{-1}$, respectively

\begin{equation}\label{eq22}
-i \mathfrak{h}_{+1,\pm1} \frac{\partial  f_{+1,\pm1}^{(1)}}{\partial \tau}-i \mathcal{H}_{+1,\pm1} \frac{\partial^{2}  f_{+1,\pm1}^{(1)}}{\partial \zeta^{2}_{+1,\pm1}}  +  a  (k  \pm Q_{+1,\pm1}) [  |f_{+1,\pm1}^{(1)}|^{2} + 2 |f_{+1,\mp1}^ {(1)}|^{2} ]f_{+1,\pm1}^{(1)}=0,
         $$$$
+i \mathfrak{h}_{-1,\pm1}\frac{\partial  f_{-1,\pm1}^{(1)}}{\partial \tau}+ i\mathcal{ H}_{-1,\pm1} \frac{\partial^{2}  f_{-1,\pm1}^{(1)}}{\partial \zeta^{2}_{-1,\pm1}}  + a   ( k \mp Q_{-1,\pm1})(| f_{-1,\pm1}^ {(1)}|^{2}  + 2  |f_{-1,\mp1}^ {(1)}|^{2}  )f_{-1,\pm1}^{(1)}=0.
\end{equation}

We consider the equations proportional to the $Z_{+1}$ in detail, the complex-conjugation equation proportional to the $Z_{-1}$ can be considered similarly.

After transformation back to the variables $z$ and $t$, from the Eq.(14) we obtain the coupled NSEs in the form
\begin{equation}\label{cnse}
i (\frac{\partial \lambda_{\pm}}{\partial t}+ v_{\pm}\frac{\partial  \lambda_{\pm}} {\partial z})+p_{\pm} \frac{\partial^{2} \lambda_{\pm} }{\partial z^{2}} + \mathfrak{q}_{\pm} (  |\lambda_{\pm}|^{2} + 2 |\lambda_{\mp}|^{2} )\lambda_{\pm}=0,
\end{equation}
where
\begin{equation}\label{ppe}
\lambda_{\pm}=\varepsilon  f_{+1,\pm1}^{(1)},
$$$$
p_{\pm}=\beta\frac{2 (   k   \pm   Q_{\pm1})  v_{\pm} +  ( \omega \pm \Omega_{\pm1})}{  1-\beta ( k\pm Q_{\pm1} )^{2}},
$$$$
\mathfrak{q}_{\pm}= - \frac{ a   (k \pm Q_{\pm 1})}{1-\beta ( k\pm Q_{\pm1} )^{2}},
$$$$
 v_{\pm}=\frac{ C + 2\beta  (\omega   \pm \Omega_{\pm1})(k \pm  Q_{\pm1})}{1 - \beta( k\pm   Q_{\pm1})^{2} },
 $$$$
\Omega_{+1,+1}=\Omega_{-1,-1}=\Omega_{+1},\;\;\;\;\;\;\;\;\;\;\;\Omega_{+1,-1}=\Omega_{-1,+1}=\Omega_{-1},
$$$$
Q_{+1,+1}=Q_{-1,-1}=Q_{+1},\;\;\;\;\;\;\;\;\;\;\;Q_{+1,-1}=Q_{-1,+1}=Q_{-1}.
\end{equation}

The solution of Eq.(15) is given by Eq.(18). Substituting Eq.(18) into the Eqs.(6) and (2) we obtain the two-component vector breather OSDFW of the modified BBM equation Eq.(1) in the form
\begin{equation}\label{myi}
u(z,t)=
\frac{2}{\mathfrak{b} T}sech(\frac{t-\frac{z}{V_{0}}}{T})\{  K_{+} \cos[(k+Q_{+1}+k_{+})z -(\om +\Omega_{+1}+\omega_{+}) t]+
$$$$
K_{-}\cos[(k-Q_{-1}+k_{-})z -(\om -\Omega_{-1}+\omega_{-})t]\}.
\end{equation}
where T is the width of the two-component nonlinear pulse (see, Appendix).

\vskip+0.5cm
\section{Conclusion }

Using generalized PRM the two-component vector breather OSDFW  solution of the modified BBM equation (1) is obtained.
The first term of Eq.(17) is the small amplitude breather oscillating with the sum  of the frequencies $\om +\Omega_{+1}$ and wavenumbers $k+Q_{+1} $(taking into account Eqs.(19)) and the second term is the small amplitude breather oscillating with the difference of the frequencies $\om -\Omega_{-1}$ and wavenumbers $k-Q_{-1} $. The nonlinear connection between breathers are determined by the terms $|\lambda_{-}|^{2} \lambda_{+}$ and $|\lambda_{+}|^{2} \lambda_{-}$ of Eq.(15). The parameters of the nonlinear wave by the equations (8), (11), (16), (20), and (21) are determined. The dispersion relation and connections between oscillating parameters are presented by  Eqs.(4), (12) and (13).

When comparing PRM and the generalized version of PRM we can see the following differences.

The PRM  adapted for the study of single-component solitary waves, and uses one complex auxiliary function and two constant parameters. As a result,  the considered  nonlinear differential equation is reduced to the scalar NSE for the single complex auxiliary function that has a soliton solution.

In contrast to the PRM,  the generalized PRM uses the two complex auxiliary functions and eight constant parameters, that  to give possibility  to study two-component nonlinear solitary waves. By means of this method  the considered nonlinear differential equation is transformed  to the coupled NSEs for the two complex auxiliary function having the vector soliton solution Eq.(18).

Using the generalized PRM, it became possible to find the solutions of  the series of different nonlinear differential equations (see, above) and obtain for these equations the solutions in the form of the two-component vector breathers  OSDFW.
These circumstances  give grounds to hope that the two-component vector breathers OSDFW, also can be found for other nonlinear equations.

\vskip+0.5cm
\section{Appendix }

The vector soliton (VS) solution of Eq.(15) can be written as (see, for instance [14-16, 28, 33] and references therein)
\begin{equation}\label{ue1}
\lambda_{\pm}=\frac{K_{\pm }}{\mathfrak{b} T}Sech(\frac{t-\frac{z}{V_{0}}}{T}) e^{i(k_{\pm } z - \omega_{\pm } t )},
\end{equation}
where $K_{\pm },\; k_{\pm }$ and $\omega_{\pm }$ are the real constants, $V_{0}$ -the velocity of the nonlinear wave. We assume that
\begin{equation}\label{kom}
k_{\pm }<<Q_{\pm 1},\;\;\;\;\;\;\omega_{\pm }<<\Omega_{\pm 1}.
\end{equation}

The parameters of VS Eq.(18) are given by
\begin{equation}\label{rrw}
T^{-2}=V_{0}^{2}\frac{v_{+}k_{+}+k_{+}^{2}p_{+}-\omega_{+}}{p_{+}}, \;\;\;\;\;\;\;k_{\pm }=\frac{V_{0}-v_{\pm}}{2p_{\pm}},
$$$$
\mathfrak{b}^{2}=\frac{V_{0}^{2}\mathfrak{ q}_{+}}{2p_{+}}(K_{+}^{2}+2 K_{-}^{2}) .
\end{equation}

The connections between  both components of VS are defined as
\begin{equation}\label{ttw}
K_{+}^{2}=\frac{p_{+}\mathfrak{q}_{-}- 2p_{-}\mathfrak{q}_{+}}{p_{-}\mathfrak{q}_{+}-2 p_{+}\mathfrak{q}_{-}}K_{-}^{2},
\;\;\;\;\;\;\;\;\;
\omega_{+}=\frac{p_{+}}{p_{-}}\omega_{-}+\frac{V^{2}_{0}(p_{-}^{2}-p_{+}^{2})+v_{-}^{2}p_{+}^{2}-v_{+}^{2}p_{-}^{2}
}{4p_{+}p_{-}^{2}}.
\end{equation}

When $a>0$ and $\beta <0$, from the Eq.(16) we obtain that $p_{\pm} \mathfrak{q}_{\pm}>0$ and consequently,  both components of VS are bright solitons and this case corresponds to a two-component bright VS.

\vskip+0.5cm


\begin{thebibliography}{20}
\bibitem[{\citenamefont{note}(1996}]{note2::}
  {\bibinfo{title}{As a point of clarification, there have been two different definitions of the modified BBM equation in the literature, which does confuse.
Indeed, the generalized modified BBM equation has the form $u_{ t}+ u_{z}+ u_{zzz} +a u^{n} u_{z}=0$, which when $n=2$ called the  nonlinear
dispersive modified BBM equation [48-51].
On the other hand, there is another definition of the modified BBM equation Eq.(1), which sometimes also called the modified regularized long wave equation.
Here, we are using this definition for the Eq.(1).}}
\end{thebibliography}
\end{document}